\documentclass[%
 reprint,
 amsmath,amssymb,aps,
pra,
]{revtex4-2}
\usepackage{float}
\usepackage{xcolor, soul}
\sethlcolor{green}
\usepackage{graphicx}
\usepackage{dcolumn}
\usepackage{bm}

\usepackage[utf8]{inputenc}
\usepackage[normalem]{ulem}
\usepackage[T1]{fontenc}
\usepackage[english]{babel}

\usepackage{amsmath}
\usepackage{amsfonts}

\usepackage{mdframed}

\usepackage[absolute]{textpos}
\usepackage{colortbl}
\usepackage{array}

\begin{document}
\preprint{APS/123-QED}

\title{Modulation instability in nonlinear flexible mechanical metamaterials}
\author{A.\,Demiquel}
\author{V.\,Achilleos}
\author{G.\,Theocharis}
\author{V.\,Tournat}

\affiliation{Laboratoire d’Acoustique de l’Université du Mans (LAUM), UMR 6613, Institut d’Acoustique - Graduate School (IA-GS), CNRS, Le Mans Université, France}

\date{\today}

\begin{abstract}
In this paper, we study modulation instabilities (MI) in a one-dimensional chain configuration of a flexible mechanical metamaterial (flexMM).
Using the lumped element approach, flexMMs can be modeled by a coupled system of discrete equations for the longitudinal displacements and rotations of the rigid mass units. In the long wavelength regime, and applying the multiple-scales method we derive an effective nonlinear Schrödinger equation for slowly varying envelope rotational waves. We are then able to establish a map of the occurrence of MI to the parameters of the metamaterials and the wavenumbers. We also highlight the key role of the rotation-displacement coupling between the two degrees of freedom in the manifestation of MI. All analytical findings are confirmed by numerical simulations of the full discrete and nonlinear lump problem. These results provide interesting design guidelines for nonlinear metamaterials offering either stability to high amplitude waves, or conversely being good candidates to observe instabilities. 

\end{abstract}

\maketitle


\section{Introduction}

In the context of nonlinear waves, flexible mechanical metamaterials have recently emerged as a rich and versatile platform, opening the way for fundamental studies and potential applications \cite{deng_nonlinear_2021}. Such flexible mechanical metamaterials (flexMMs) can be defined as artificial compliant structures able to support large deformations and mechanical instabilities leading to new modes of functionality~\cite{bertoldi_flexible_2017}. As a result, a plethora of original quasi-static behaviors and functions have already been reported, with applications to soft robotics~\cite{rafsanjani_programming_2019}, structure reconfigurability~\cite{haghpanah_multistable_2016} or mechanical logic devices~\cite{raney_stable_2016,bilal_bistable_2017,jiang_bifurcation-based_2019}, as examples. In addition and more recently, the study of their dynamic properties has revealed that the nonlinearity is most often geometric in nature, resulting from large local deformations, which makes the nonlinear dynamic response governed by the architecture and therefore controllable \cite{deng_nonlinear_2021}. This latter possibility opens the way to targeting specific dynamical properties, which have been known to be described by existing fundamental equations (such as nonlinear Klein-Gordon equations found in \cite{deng_elastic_2017}) or which could illustrate and reveal new relevant dynamic equations.

Up to now, the specific behaviors of the reported flexMM designs could be accurately modeled as rigid units able to translate and rotate, connected with highly compliant springs of longitudinal, shear and bending nature. On the one hand, the derived nonlinear and discrete equations of motion for multiple degrees of freedom can be efficiently solved by numerical integration \cite{deng_nonlinear_2021}. On the other hand, several steps towards analytical solutions can be taken, including the consideration of periodicity, long wavelength compared to the lattice period, expansions to first order nonlinear and dispersive terms, for instance. A review of the main nonlinear wave processes and corresponding equations in flexMM reported to date can be found in \cite{deng_nonlinear_2021}. These include among others the observation of mechanical vector solitons, their interactions and tuning \cite{deng_elastic_2017,deng_metamaterials_2018,deng_nonlinear_2021}, the observation of cnoidal waves \cite{mo_cnoidal_2019} and of transition waves \cite{haghpanah_multistable_2016,raney_stable_2016,bilal_bistable_2017,jiang_bifurcation-based_2019}.
However, nonlinear modulated waves in flexMM is an 
unexplored field. Many interesting wave phenomena are expected to be revealed,
including the manifestation of modulation instability (MI) and the resulting formation of localized waves such as envelope solitons or breathers \cite{zakharov_modulation_2009,dudley_rogue_2019,copie_physics_2020}. 


The phenomenon of MI has attracted a significant research interest in a range of different wave systems, both continuum (water surface \cite{benjamin_disintegration_1967,chabchoub_rogue_2011,chabchoub_time-reversal_2014}, plasmas \cite{ghosh_modulational_1985},  optical fibers \cite{tai_observation_1986,shukla_modulational_1986}, Bose–Einstein condensates \cite{strecker_formation_2002}) and discrete (electrical transmission lines \cite{kengne_modulational_2006}, granular chains \cite{liu_breathers_2016}) described by the universal nonlinear Schrödinger equation (NLSE) \cite{ablowitz_discrete_2004,peyrard_physique_2004,solli_optical_2007}. 
MI analysis conventionally describes the early (linear) stage of the exponential growth of perturbations of an unstable plane wave background \cite{tai_observation_1986,shukla_modulational_1986,potasek_modulation_1987,cheng_controllable_2014,diakonos_symmetric_2014,kraych_statistical_2019}. 
Recently, a renewed interest in MI has appeared, motivated by the search for extreme waves, and has led to the analysis of various initial conditions not limited to plane waves as well as to the study of the subsequent nonlinear stages of instability beyond the initial linear stage. \cite{sarma_modulational_2010,sarma_modulational_2011,xiang_modulation_2011}. Along these lines, numerous theoretical and experimental works in water wave tanks and optical fibers appeared in the literature \cite{chabchoub_rogue_2011,chabchoub_time-reversal_2014,bonnefoy_modulational_2020,xu_observation_2020,kibler_peregrine_2010,tikan_universality_2017,tikan_local_2021,pierangeli_observation_2018,liu_breathers_2016,zhao_controlled_2017,copie_physics_2020}. 

It is the main objective of this paper to study the phenomenon of MI in nonlinear flexMM. 
To do so, starting from a discrete, nonlinear lump model, which was found to describe well the dynamics of flexMM, we derive a NLS equation for the slowly varying envelope of waves of the rotational degree of freedom. Then, we analyze under which conditions, modulation instability of plane waves emerges by random perturbations. We finally compare the theoretical results with numerical simulations of the full nonlinear lump model. We show that, via an initial condition problem, the coupling between the degrees of freedom of the particles as well as the mechanical parameters of the metamaterial (see section \ref{sec:level1}), can allow modulation instability to occur and under which conditions. 


\begin{figure*}
\includegraphics[width=0.9\textwidth]{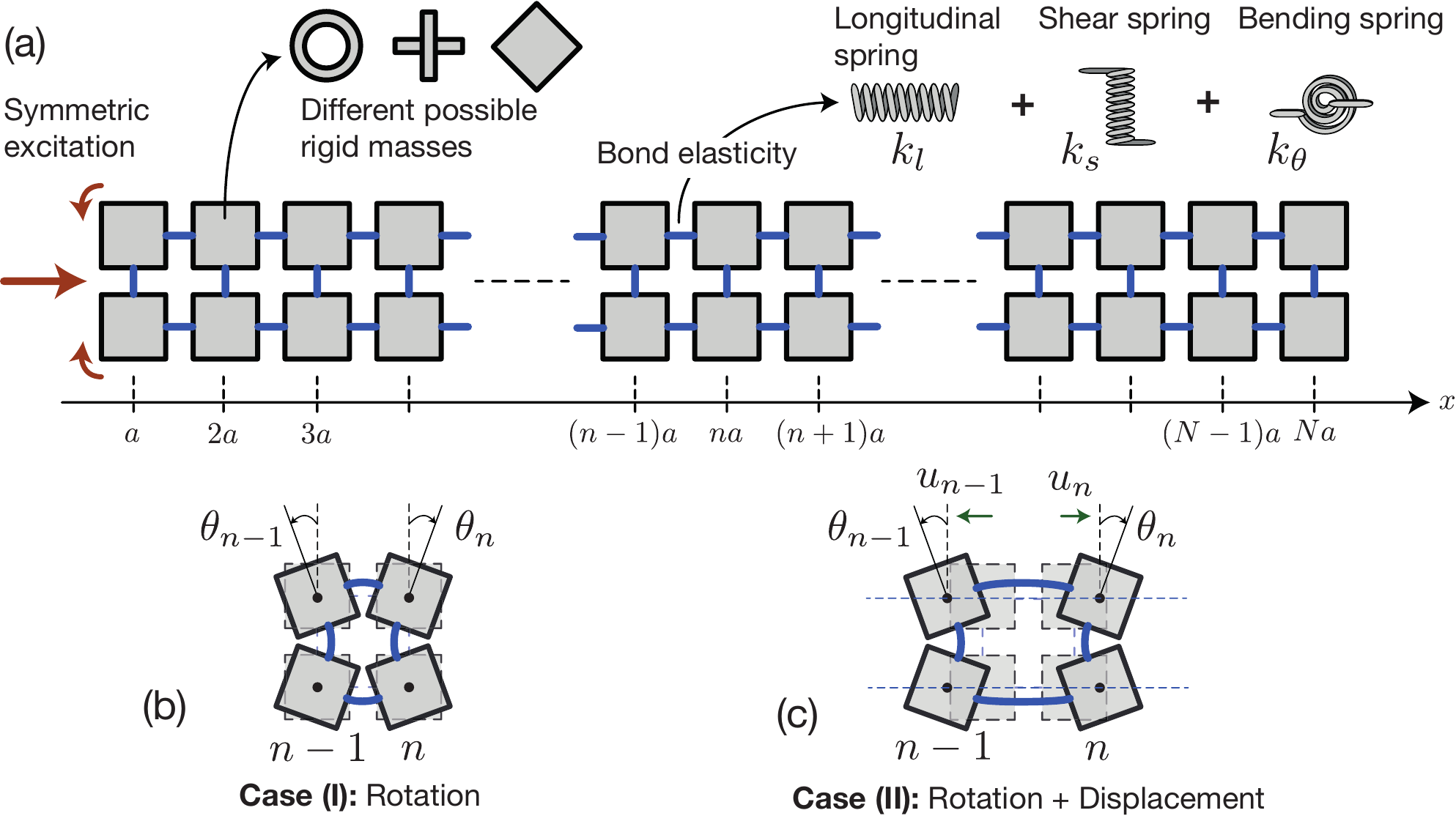}
\caption{\label{fig:}  (a) Sketch of the chain configuration periodic flexMM under consideration. It is composed by two rows of rigid mass units (gray squares) linked by elastic connectors (thick, blue lines) extending along x-direction with a lattice constant $\alpha$.
The rigid units can be of various shapes (for example crosses, spheres, cubes) and are characterized by a mass $m$ and a moment of inertia $J$. The elastic bonds (for example highly flexible plastic films) are characterized by three effective stiffness. 
We consider symmetric movements relative to the horizontal symmetry axis of the system. The displacements of the $n$ and $n-1$ particles from the equilibrium position are shown in panels (b) and (c) for the two different considered cases. In panel (b) the mass units can only rotate, case (I) while in panel (c), the mass units can both rotate and longitudinally translate, case (II).}
\end{figure*}

\section{\label{sec:level1}Properties and modeling of the considered Flexible Mechanical Metamaterial}
\subsection{\label{sec:level2} Problem position and modeling of the structure }

The considered structure is inspired from the flexible Lego chain implemented in ref.~\cite{deng_metamaterials_2018} and it consists of rigid units (an assembly of Lego bricks), that are linked to the next neighbors by highly flexible plastic films. A periodic chain can then be constructed by connecting pairs of units along one direction as shown in Fig.~\ref{fig:}(a). The plastic films connecting the rigid bodies are physically modeled by massless springs. Three springs are needed to represent the plastic films connections, a longitudinal spring with stiffness $k_l$, a shear spring with a shear stiffness $k_s$, and a bending spring with a bending stiffness $k_{\theta}$. Two rows of masses were originally used in ref.~\cite{deng_metamaterials_2018} because this chain configuration possesses a symmetry axis ensuring symmetry of the motion and no experimental buckling of the chain out of this axis. The motion takes place in the plane of the chain and in the general case, each mass should have 3 degrees of freedom, one rotation and two displacements. In the context of soliton propagation \cite{deng_metamaterials_2018,deng_elastic_2017}, it has been shown numerically and experimentally that ignoring the transversal displacement is a reasonable assumption. Indeed, the numerically and experimentally observed transversal displacement amplitude is an order of magnitude smaller than the longitudinal one. A 2-degree-of-freedom (dof) model was therefore used for this system, and could be used as a starting point for obtaining relevant analytical solutions. 

In the present study, we also ignore the transversal displacements and we consider two cases. Case (I), Fig.~\ref{fig:}(b), where each rigid unit is free only to rotate (thus is described by one dof $\theta$), and case (II), Fig.~\ref{fig:}(c), where each rigid unit both rotates and is longitudinally displaced (thus is described by two dofs $\theta$ and $u$).  
Based on the mirror symmetry of the two lines configuration along the y-axis, we look for symmetric excitations for which the two rigid units of each column move along $x$ with the same amount and rotate at an opposite angle. 

As done in \cite{deng_metamaterials_2018}, a positive direction of rotation is from now on defined alternately for neighboring units since the natural rotation is alternated, upon static compression or long-wavelength propagation. The corresponding normalized equations of motion for the $n$-th column are then written \cite{deng_metamaterials_2018},

\begin{equation}
\begin{aligned}
\frac{\partial^{2} U_{n}}{\partial T^{2}} &= U_{n+1}-2 U_{n}+U_{n-1}-\frac{\cos \theta_{n+1}-\cos \theta_{n-1}}{2}, \\
\frac{1}{\alpha^2} \frac{\partial^{2} \theta_{n}}{\partial T^{2}} &= -K_{\theta}\left(\theta_{n+1}+4 \theta_{n}+\theta_{n-1}\right)\\
&+K_{s} \cos \theta_{n}\left[\sin \theta_{n+1}+\sin \theta_{n-1}-2 \sin \theta_{n}\right] \\
&-\sin \theta_{n}\left[2 \left(U_{n+1}-U_{n-1}\right)+4-2 \cos \theta_{n}\right.\\
&\left.-\cos \theta_{n+1}-\cos \theta_{n-1}\right] ,
\end{aligned}
\label{norm}
\end{equation}
where we have introduced the following normalized variables and parameters: the longitudinal displacement of unit $n$,  $U_n=u_n/a$, the normalized time $T=t\sqrt{k_l/m}$, an inertial parameter $\alpha =a/\sqrt{m/(4J)}$, and stiffness parameters $K_{\theta} = 4 k_{\theta}/k_l a^2$ and $K_s = k_s/k_l$. Above, $m$ and $J$ are the  mass and the moment of inertia of the rigid units, while $a$ is the unit cell length (distance between the centers of the masses).

\subsection{Discrete dispersion relations}

A particularity of this system, compared to other mechanical chains with two dofs, \cite{pichard_localized_2014,prodan_dynamical_2017,kopfler_topologically_2019,allein_linear_2020,miyazawa_topological_2022}, is that in the linear limit, 
the two motion (displacements and rotations) are decoupled, i.e. each degree of freedom follows its own dynamics, independent of the other (see Appendix).  

The corresponding dispersion relations are given by
    \begin{align}
        \omega^{(1)}&= 2\sin \left(\frac{qa}{2}\right)\label{discrete1} \, , \\
    \omega^{(2)} &=\pm \sqrt{4\alpha^2(K_s-K_{\theta})\sin^2\left(\frac{qa}{2}\right)+6\alpha^2K_\theta}\, .
        \label{discrete2}
    \end{align}

The first branch, Eq.~(\ref{discrete1}), describes propagating longitudinal waves with the typical monoatomic dispersion relation.
The second branch, Eq.~(\ref{discrete2}), describes propagating rotational waves with a Klein-Gordon type dispersion relation and a lower cutoff frequency at $\omega=\alpha\sqrt{6K_\theta}$. From Eq.~(\ref{discrete2}), it is clear that the dispersion relation of the structure can be highly tuned through the inertial parameter $\alpha$ (changing the mass and the shape of the rigid particles) as well as the stiffness parameters $K_s$, $K_{\theta}$ (changing the elastic parameters of the plastic films). Four examples of the dispersion relation for different values of the bending stiffness $K_{\theta}$ are shown in Fig.~\ref{Dispersion relations} with solid lines. The rest of the parameters are chosen  to be consistent with the literature \cite{deng_elastic_2017,deng_metamaterials_2018,guo_nonlinear_2018,mo_cnoidal_2019}. Note also that the concavity of the dispersion relation for the rotation dof is defined by the sign of $\delta = K_s-K_\theta$, see Fig.~\ref{Dispersion relations}(a-c) vs (b-d). As we explain below, the sign of $\delta$ plays a key role in the stability of the plane waves in the system.

\subsection{Continuum Limit}
 
Considering waves with wavelengths that are sufficiently larger than the unit cell distance, i.e. $\lambda \gg a$, 
one can employ the continuum limit approximation. 
%
Therefore, we define two continuous functions $U(X,T)$ and $\theta(X,T)$, interpolating the displacement and rotation of the $n$-th pair of rigid units located at the position $x_n = na$, where $n$ is an integer, such that 

\begin{equation}
    U(X_n,T)=U_n(T)\, , \;\;  \;\; \theta(X_n,T)= \theta_n(T)\, ,\;\; X_n =\frac{x_n}{a}\, .
\end{equation}
%
If we further assume weak nonlinearity, namely $\theta\ll 1$, keeping terms up to $\theta^3$, see also \cite{deng_metamaterials_2018,deng_elastic_2017}, Eqs.~(\ref{norm}) yield,
\begin{align}
    \frac{\partial^2U}{\partial T^2} &=  \frac{\partial^2U}{\partial X^2}+\theta \frac{\partial \theta}{\partial X} \label{eqnorm1}\, ,\\
    \frac{\partial^{2} \theta}{\partial T^{2}}&= C_1 \frac{\partial^2\theta}{\partial X^2}-C_2\theta-C_3\theta^3-C_4\theta \frac{\partial U}{\partial X}\, ,
    \label{eqnorm2}
\end{align}
%
%
where $C_1 = \alpha^2[K_s-K_{\theta}]$, $C_2 = 6K_{\theta}\alpha^2$, $C_3 = 2\alpha^2$ and $C_4 = 4\alpha^2$. The system of equations (\ref{eqnorm1}-\ref{eqnorm2}) is  a simple dispersion-less wave equation for the displacement field $U$, Eq.~(\ref{eqnorm1}),  coupled through a nonlinear term, with a Klein-Gordon equation for the rotation field $\theta$, Eq.~(\ref{eqnorm2}). 
Pulse soliton solutions of (\ref{eqnorm1}-\ref{eqnorm2}) were theoretically obtained and experimentally observed in \cite{deng_metamaterials_2018}, revealing the validity of the continuum coupled equations.

%



\begin{figure}[ht!]
\begin{center}
\includegraphics[width=0.49\textwidth]{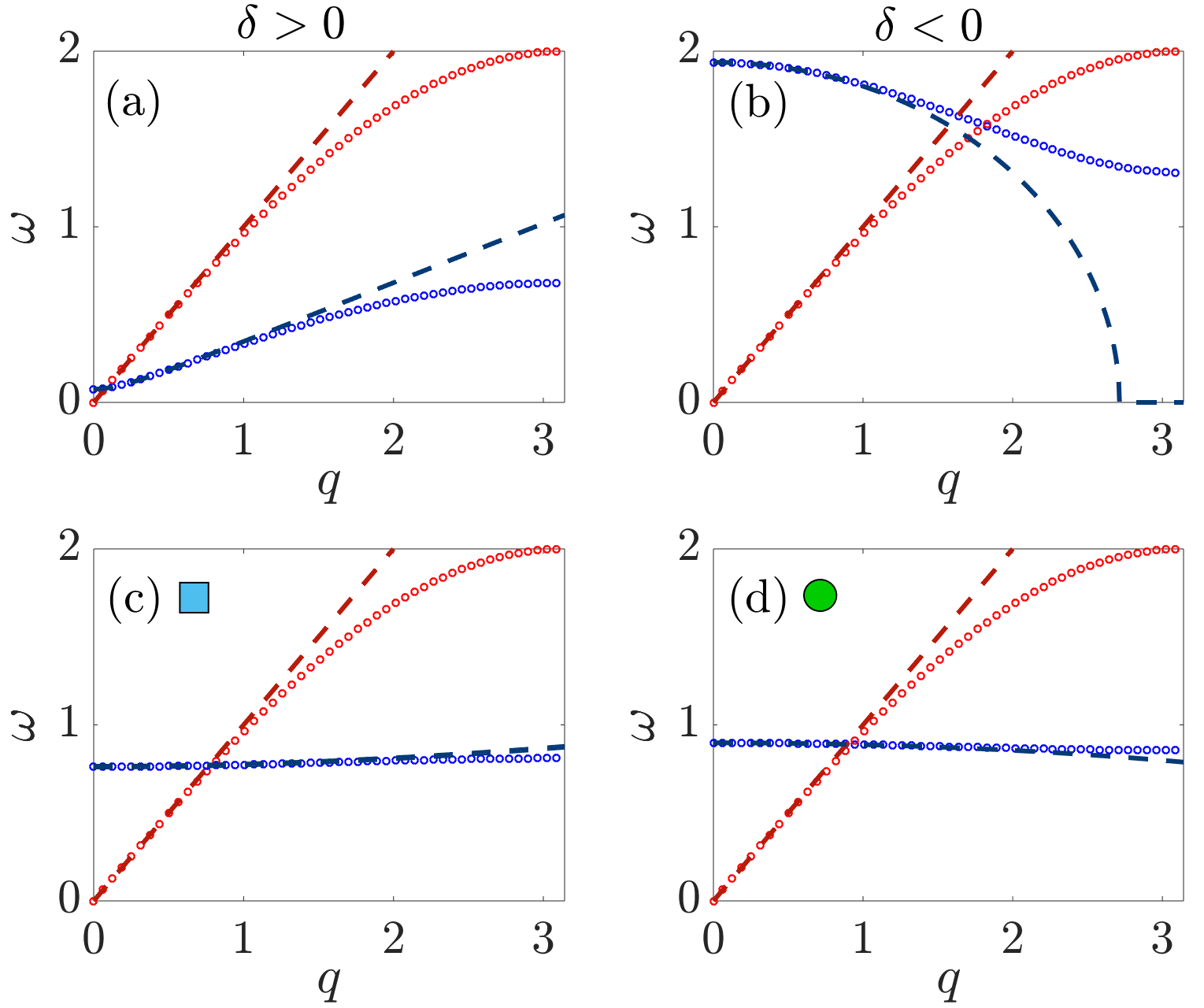}
	\caption{\label{Dispersion relations} Dispersion relations of Eqs.~(\ref{discrete1}-\ref{discrete2}) (solid lines) and of the continuum approximation (dashed lines) following Eqs.~(\ref{continuum1}-\ref{continuum2}). 
	In all the examples, we fix the coefficients $\alpha= 2.5$, $K_s = 0.01851$ and we vary $K_{\theta}$. (a) $K_{\theta} =1.534.10^{-4}$, (b) $K_{\theta}=0.1$, (c) $K_{\theta}=0.01551$, and (d) $K_{\theta}=0.02151$. Left (right) panels correspond to $\delta>0$ ($\delta <0)$.}
	\vspace{-0.6cm}
	\label{disp_rel_example}
\end{center}
\end{figure}

The linear dispersion relations of Eqs.~(\ref{eqnorm1}-\ref{eqnorm2}) are given by:
    \begin{align}
        \omega^{(1)}&= k\label{continuum1} \, ,\\
    \omega^{(2)}&= \sqrt{C_1 k^2 + C_2}\, ,
        \label{continuum2}
    \end{align}
and they are shown in Fig.~\ref{Dispersion relations} with dashed lines.
For the cases we plot, one can see that as long as the wavenumber $q\leq 1$, the continuum equations capture well the dispersive characteristics of the discrete model.

\section{Modulated waves in flexMM}

Although there are several recent studies on pulse nonlinear waves, the existence, stability and propagation of nonlinear modulated waves, in the form of plane waves or wavepackets in flexMM remain unexplored. Only recently, the existence and stability of discrete breathers in flexMM was explored \cite{duran_discrete_2022}.
Here, we derive the theoretical framework for the description of long-wave, nonlinear modulated waves. For this purpose, we apply below the multiple-scales method  \cite{peyrard_physique_2004,holmes_introduction_1995} to Eqs.~(\ref{eqnorm1}-\ref{eqnorm2}).

\subsection{Multiple-Scales}
We are looking for $U$ and $\theta$, in the form of a perturbative expansion,
\begin{equation}
    \begin{aligned}
    U &= \sum_{i=1}^{N}\epsilon^i u_i(X_0,\ldots X_N,T_0,\ldots T_N)\, ,\\
     \theta  &= \sum_{i=1}^{N}\epsilon^i \theta_i(X_0,\ldots X_N,T_0,\ldots T_N)\, ,\\
    \end{aligned}
    \label{perturbatif}
\end{equation}
where $T_i =\epsilon^i T$ and $X_i =\epsilon^i X$, with $i=0,1,\ldots N$ and $\epsilon$ represents a small parameter. $X_0$ and $T_0$ correspond to the original “fast” spatial and temporal scales of the carrier wave, while $X_i$ and $T_i$ with $i\neq 0$ define progressively the “slow” spatial and temporal scales of the envelope.

By inserting the expansions of  Eq.~(\ref{perturbatif}) into the system of Eqs.~(\ref{eqnorm1}-\ref{eqnorm2}), and taking into account the derivative operators of the new spatial and temporal variables [see appendix \ref{appendix:Multiplescales} Eqs.~(\ref{diff})], we end up with the following hierarchy of equations at successive orders of $\epsilon$,
\begin{equation}
    \begin{aligned}
 &\mathcal{O}(\epsilon) \\  &
    \begin{cases}
        &\hspace*{-0.35cm}\mathcal{\hat{L}}_0^{(1)}u_1 = 0 \, ,\\
        &\hspace*{-0.35cm}\mathcal{\hat{L}}_0^{(2)}\theta_1 = 0 \, ,
    \end{cases}
    \\
    \\
 &\mathcal{O}(\epsilon^2)\\ & 
        \begin{cases}
        &\hspace*{-0.35cm}\mathcal{\hat{L}}_0^{(1)}u_2= -\mathcal{\hat{L}}_1^{(1)}u_1+\mathcal{\hat{M}}_0^{(1)}\theta_1^2 \, ,\\
        &\hspace*{-0.35cm}\mathcal{\hat{L}}_0^{(2)} \theta_2 = -\mathcal{\hat{L}}_1^{(2)}\theta_1 + \theta_1 \mathcal{\hat{M}}_0^{(2)}u_1 \, ,\\
        \end{cases}
        \\
        \\
  &\mathcal{O}(\epsilon^3) \\  &
    \begin{cases}
        &\hspace*{-0.35cm}\mathcal{\hat{L}}_0^{(1)}u_3 = -\mathcal{\hat{L}}_1^{(1)}u_2-\mathcal{\hat{L}}_2^{(1)}u_1+ \mathcal{\hat{M}}_1^{(1)}\theta_1^2+2\mathcal{\hat{M}}_0^{(1)}\theta_1\theta_2 \, , \\
        &\hspace*{-0.35cm}\mathcal{\hat{L}}_0^{(2)}\theta_3  = -\mathcal{\hat{L}}_1^{(2)}\theta_2-\mathcal{\hat{L}}_2^{(2)}\theta_1+\mathcal{\hat{M}}^{(3)} \theta_1^3+\theta_1\mathcal{\hat{M}}_0^{(2)}u_2\\
        &\hspace{1.4cm}+\theta_1\mathcal{\hat{M}}_1^{(2)}u_1+\theta_2\mathcal{\hat{M}}_0^{(2)}u_1 \, ,
     \end{cases}
    \end{aligned}
    \label{msm}
\end{equation}
where the linear operators, $\mathcal{\hat{L}}_j^{(i)}$ and $\mathcal{\hat{M}}_j^{(i)}$, applied to the linear and nonlinear terms of the equations (\ref{eqnorm1}-\ref{eqnorm2}) respectively are defined in the appendix \ref{appendix:Multiplescales} Eqs.~(\ref{operators}). 

The first set of equations (\ref{msm}) of order $\mathcal{O}(\epsilon)$, corresponds to the linearized system of 
Eqs.(\ref{eqnorm1}-\ref{eqnorm2}). Using the fact that in the linear regime the two fields are decoupled, we will focus on the particular case which, at the leading order, there is only rotational motion, i.e., 
\begin{equation}
\begin{aligned}
    u_1 &= 0 \, , \\
    \theta_1 &= B(X_1,T_1,X_2,T_2,...)e^{i(k X_0-\omega T_0)} +  \text{c.c}\, ,
\label{leadingorder}
\end{aligned}
\end{equation}
with $\omega$ and k satisfying the dispersion relation Eq.\,(\ref{continuum2})
%
and c.c stands for the complex conjugate. 

Let us proceed to the next order of the perturbation scheme, $\mathcal{O}(\epsilon^2)$, and substitute the solutions (\ref{leadingorder}) into the second set of equations (\ref{msm}) to obtain,

\begin{equation}
\begin{split}
       \mathcal{\hat{L}}_0^{(1)}u_2&=\mathcal{\hat{M}}_0^{(1)}\theta_1^2 \, ,\\
        \mathcal{\hat{L}}_0^{(2)} \theta_2& = -\mathcal{\hat{L}}_1^{(2)}\theta_1 \, .\\
        \end{split}
\end{equation}

The right-hand-side of the last equation is a secular term, as it acts as a source term proportional to $e^{i\sigma}$ ($\sigma = k X_0-\omega T_0$) with which the linear operator $\mathcal{\hat{L}}_0^{(2)}$ on the left
is in resonance. This implies that the solution $\theta_2$ would blow up as $t\rightarrow \infty$ and thus the perturbation scheme will fail. The only way for the expansion to be bounded is to set the secular term to zero,  
which translates to the following relation for the envelope function $B$,
\begin{equation}
 D_1 B + v_g D_{1X}B =0 \,.
    \label{solvability_condition}
\end{equation}
Here we have introduced the group velocity given by 
\begin{equation}
    v_g =\frac{C_1 k}{\sqrt{C_1 k^2+C_2}} = \frac{C_1 k}{\omega}  \,.
\end{equation}
Once the secular term is removed,
the system of equations of the second order in $\epsilon$ in Eq.\,(\ref{msm}) is now reduced to,
\begin{equation}
         \begin{cases}
        &\mathcal{\hat{L}}_0^{(1)}u_2=   ik B^2e^{2i\sigma}+ \text{c.c} \, ,\\
        &\mathcal{\hat{L}}_0^{(2)} \theta_2 = 0 \, .\\
        \end{cases}
\end{equation}

The first equation has the following solution, 
\begin{equation}
    u_2 = \frac{ikB^2}{4(k^2-\omega^2)}e^{2i\sigma}+ \text{c.c} \, ,
    \label{u1}
\end{equation}
where the homogeneous part of the solutions is omitted due to our choice of initial conditions $U(0,X)=\dot{U}(0,X)=0$.
For $\theta_2$ we choose the trivial solution, i.e. $\theta_2 = 0$, since any other solution can be incorporated in $B$. 

\subsection{Nonlinear Schrödinger Equation (NLSE)}

We now proceed with the $\mathcal{O}(\epsilon^3)$ order of the perturbation scheme.
By using $u_1=0$ and $\theta_2=0$, as discussed above, the last equation of 
Eq.~(\ref{msm}) is reduced to
\begin{equation}
    \mathcal{\hat{L}}_0^{(2)}\theta_3  = -\mathcal{\hat{L}}_2^{(2)}\theta_1+\mathcal{\hat{M}}^{(3)} \theta_1^3+\theta_1\mathcal{\hat{M}}_0^{(2)}u_2 \, .
    \label{Order_3}
\end{equation}
Similar to the previous order, there are secular terms in the right-hand side of Eq.~(\ref{Order_3})
proportional to $e^{i\sigma}$: the $\mathcal{\hat{L}}_2^{(2)}\theta_1$, and parts of the 
$\mathcal{\hat{M}}^{(3)}\theta_1^3$ and $\theta_1\mathcal{\hat{M}}_0^{(2)}u_2$ terms.
To find their secular contributions, we develop the operators as well as the functions on which they are applied. For the first of them,
\begin{equation}
\begin{split}
    \mathcal{\hat{M}}^{(3)}\theta_1^3 
     &= -C_3B^3e^{3i\sigma}-3C_3 \vert B\vert^2 B e^{i\sigma}+\text{c.c} \, ,
    \end{split}
    \label{secular_1}
\end{equation}
the secular contribution is $-3C_3 \vert B\vert^2 B e^{i \sigma}$.
For the next one,
\begin{equation}
\begin{split}
    \theta_1\mathcal{\hat{M}}_0^{(2)}u_2=
       \frac{C_4k^2 B^3}{2(k^2-\omega^2)} e^{3i\sigma}+\frac{C_4k^2 \vert B\vert^2 B}{2(k^2-\omega^2)} e^{i\sigma}+\text{c.c} \, ,
\end{split}
\label{secular_2}
\end{equation}
the secular contribution is $\frac{C_4k^2 \vert B\vert^2 B }{2 \left(k^2-\omega^2\right)}e^{i\sigma}$.
To avoid the resonant driving we set all the secular terms equal to zero 
(\ref{Order_3}-\ref{secular_1}-\ref{secular_2}),
 \begin{equation}
    \mathcal{\hat{L}}_2^{(2)}\theta_1+\left(3C_3 -\frac{C_4k^2}{2 \left(k^2-\omega^2\right)}\right)\vert B\vert^2 B e^{i\sigma} =0 \, .
    \label{NLS_XT_operator}
\end{equation}
It is possible to simplify this expression Eq.~(\ref{NLS_XT_operator}) by using the variables 
$\xi_i=X_i-v_gT_i$, $\tau_i=T_i$, i.e. a reference frame moving with the group velocity. Within this frame 
Eq.~(\ref{solvability_condition}), becomes $\partial B/\partial \tau_1=0$
%
and Eq.~(\ref{NLS_XT_operator}) leads to the following  nonlinear Schrödinger (NLS) equation, 
\begin{equation}
\
    i \frac{\partial B}{\partial \tau_2}+\frac{g_1}{2}\frac{\partial^2 B}{\partial \xi_1^2}+g_2 \vert B \vert ^2 B  = 0 \, .
\label{NLS}
\end{equation}
Eq.~(\ref{NLS}) describes the evolution of the envelope B of the modulated rotational waves, in the co-moving space variable and the second order slow time. 

The coefficients $g_1$ and $g_2$ are given by the following expressions,
\begin{equation}
 \begin{split}
    g_1 &= \frac{d^2\omega}{dk^2}=\frac{C_1-v_g^2}{\sqrt{C_1 k^2+C_2}} \, , \\
    g_2 &= -\frac{1}{2\sqrt{C_1 k^2+C_2}}\left(3C_3+\frac{C_4k^2}{2k^2(C_1-1)+2C_2}\right) \,. 
    \label{g_i}
\end{split}
\end{equation}
Furthermore, Eq.~(\ref{NLS}) can be rewritten as a function of a single nonlinear parameter $g =g_2/g_1$ by applying the following change of variable $\tilde{\tau}_2 = g_1 \tau_2$,
\begin{equation}
\
    i \frac{\partial B}{\partial \tilde{\tau}_2}+\frac{1}{2}\frac{\partial^2 B}{\partial \xi_1^2}+g \vert B \vert ^2 B  = 0 \, .
\label{NLS_final}
\end{equation}
In its current form, the NLS equation has two distinct behaviors depending on the sign of the nonlinearity coefficient: 
it is known as focusing when $g>0$ and defocusing for $g<0$. Among other different properties between these two cases, an important one is the stability of plane wave solutions. More precisely, for the focusing case, it is known that plane waves are subject to modulational instabilities \cite{copie_physics_2020,akhmediev_rogue_2009,akhmediev_modulation_1986,akhmediev_exact_1987,joseph_modulational_2021,cheng_controllable_2014,diakonos_symmetric_2014}, which is the  main interest of the present work. Therefore, below we establish the conditions under which MI appears in the proposed flexMM.


\subsection{Modulation instability (MI) }
We seek solutions of Eq.\,(\ref{NLS_final}) in the form of a perturbed plane wave 
\cite{zakharov_modulation_2009},
\begin{equation}
    B(\xi_1,\tilde{\tau}_2) = (A_0+  b(\xi_1,\tilde{\tau}_2))e^{i(k_0\xi_1-\omega_0 \tilde{\tau}_2+  \tilde {\theta}(\xi_1,\tilde{\tau}_2))} \, ,
    \label{perturbed}
\end{equation}
with $b$ the amplitude and $\tilde{\theta}$ the phase of small perturbations. The unperturbed plane wave satisfies the dispersion relation,
\begin{equation}
    \omega_0 = \frac{k_0^2}{2}-gA_0^2 . 
\end{equation}
Inserting Eq.~(\ref{perturbed}) into Eq.~(\ref{NLS_final}), we find at first order a set of linear equations for the perturbations $b$ and $ \tilde{\theta}$. We thus assume harmonic solutions of the form,
\begin{equation}
    \begin{split}
        b = f_1e^{i(K\xi_1-\Omega\tilde{\tau_2})}\, ,\quad 
        \tilde \theta = f_2e^{i(K\xi_1-\Omega\tilde{\tau_2})}\, , 
    \end{split}
    \label{b and theta}
\end{equation}
where the perturbation frequency $\Omega$ and wavenumber $K$ follow the dispersion relation,
\begin{equation}
    \Omega=Kk_0\pm |K|\sqrt{\frac{K^2}{4}-gA_0^2}\,.
\end{equation}
We can now identify two different regions of stability of the plane waves. On the one hand,  where $g<0$ the perturbations are oscillating functions and remain bounded. Thus we call this region \textit{modulational stable}. On the other hand, for $g>0$ there exists a band of unstable wavenumbers satisfying $K<K_c$ where,
\begin{equation}
    |K_c|=2 A_0\sqrt{g}\, ,\label{kc}
\end{equation}
resulting in a complex frequency
$\Omega=\Omega_R\pm i \Omega_I$ with
\begin{equation}
    \Omega_R=Kk_0,\quad \Omega_I=|K|A_0\sqrt{g-\frac{K^2}{4A_0^2}}\, .
    \label{Omega_I}
\end{equation}
We call this region modulational \textit{unstable}.
The small unstable wavenumbers lead to an exponential growth of the perturbations, with a growth rate $\Omega_I$. Thus any perturbation with wavenumbers within the instability band should lead to MI. 
Another important parameter for studying MI is the wavenumber with the maximum growth rate,
\begin{equation}
    |K_m|= A_0\sqrt{2g}\, .
    \label{K_m}
\end{equation}
We notice that both the critical wavenumber $K_c$ and the wavenumber corresponding to the fastest growth rate of the perturbations $K_m$, depend on the parameter $g$ and the initial amplitude $A_0$. 

\subsubsection*{Parametric study of the coefficient $g$} 

It is now clear that the stability of modulated waves in the flexMM depends on the sign $g$.
As already discussed in section \ref{sec:level1}, we study two distinct cases: (I) allowing only rotations and (II) with $2$ dofs per unit, i.e. including both rotation and longitudinal displacement [Fig.\ref{fig:}(b-c)].
The corresponding nonlinear coefficient $g(\delta,\alpha,K_\theta,k)$ for the two cases is given by,
\begin{equation}
    g=\frac{-3\alpha^2}{\delta\alpha^2-v_g^2}\label{g1}\, ,
\end{equation} for case (I), and
\begin{align}
     g=\frac{-3\alpha^2}{\delta\alpha^2-v_g^2}\left(1+\frac{k^2}{3k^2\left(\alpha^2 \delta -1\right)+18K_\theta \alpha^2}\right)\label{g2}\, ,
\end{align}
for case (II). 

\begin{figure}[h!]
   \centering
   \includegraphics[width=0.49\textwidth,trim = 0 0 0 0]{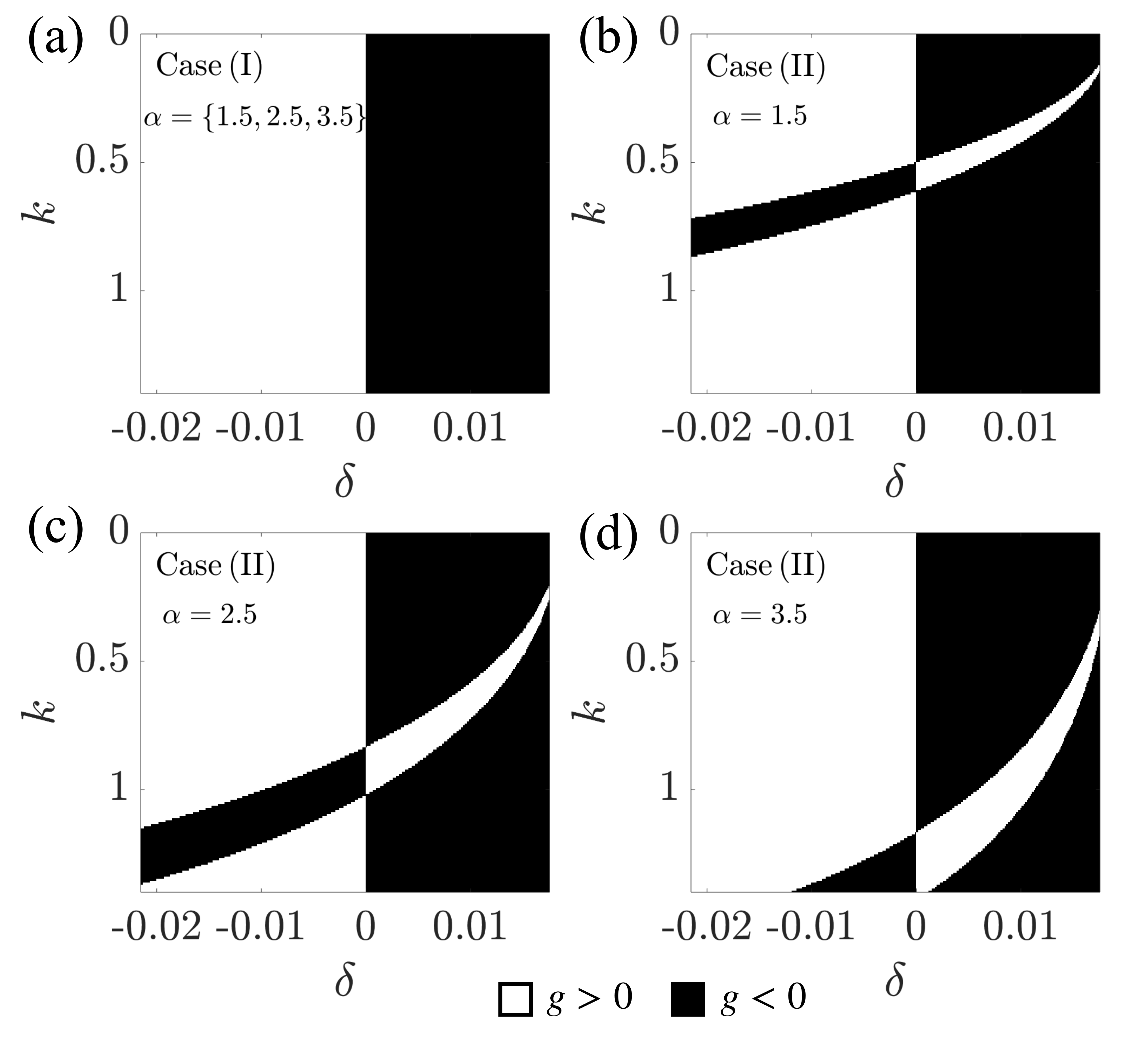}
  	\caption{Sign of the nonlinear coefficient $g$ as a function of $k$ and $\delta$, for $\alpha=1.5, 2.5, 3.5$. Panel (a) corresponds to the case (I) while panels (b,c,d) to case (II). 
  	}
  \label{Panels_delta_map}
\end{figure}

In practice, the sign of $g$ is determined by the choice of the carrier wavenumber $k$ and the geometrical characteristics of the flexMM. This shows the great flexibility that the proposed system offers in order to manipulate weakly nonlinear waves. 
In Fig.~\ref{Panels_delta_map} we plot a map of the sign of $g$ as a function of the wavenumber $k$ and $\delta$. In all cases, white (resp. black) regions correspond to $g>0$ (resp. $g<0$). From the left panel, it is clear that for case (I) with only rotations, the sign of $g$ solely depends on the sign of delta. However for case (II), things are different and the coupling between the rotation and the longitudinal motion creates intermediate regions of focusing and defocusing behavior depending also on the wavenumber $k$. The different panels of Fig.~\ref{Panels_delta_map} also show how these regions "move" towards larger $k$ by changing the value of the inertia parameter $\alpha$.

Another interpretation of the results plotted in Fig.~\ref{Panels_delta_map} is that the coupling between the rotations and longitudinal displacements creates stripes of stability (black shaded regions) in the otherwise unstable single dof lattice with only rotations [panel (a)]. At the same time this coupling forms regions of instability (white) where solely rotational motion would have been stable. Once again, this result shows the great tunability and richness of the system regarding nonlinear wave propagation.

\section{Numerical simulations of the flexMM}
\label{Discrete modeling of the FlexMM}

In this section we use direct numerical simulations of the system's discrete equations (\ref{norm}), in order to verify our analytical predictions. 
In particular we first want to check the stability of plane waves as this is predicted by the sign of $g$ (defocusing vs focusing) of the effective NLS. In addition, in the case of modulational instability, we want to compare the unstable generated wavenumber, according to the ones that the MI analysis predicts. Furthermore, we use the numerical simulations to uncover as well the dynamics of the system long after the emergence of the MI. 
We thus solve the Eqs.~(\ref{norm}) using a $4th$ order Runge-Kutta iterative integration scheme for a total of $N=500$ sites, using periodic boundary conditions. We focus on the case with $\alpha=2.5$ (Fig.~\ref{Panels_delta_map} panels (a) and (c)) although any other choice of $\alpha$ could have been done in principle. 

As initial conditions, we apply {\it plane waves} on the rotations only, with wave-number $k$, whose amplitude is perturbed by a random noise 
\begin{equation}
\begin{split}
    \theta(n,0) &= 2\epsilon (1+b_0)\cos(kn)\, ,\\
     \dot{\theta}(n,0) &= 2\epsilon\omega(k) (1+b_0)\sin(kn)\, ,
     \end{split}\label{IC}
\end{equation}
with $\epsilon=0.01$ and $b_0\in [-10^{-3},10^{-3}]$ is a random number taken from a uniform distribution. As mentioned above, in all the cases we use $U(n,0)=\dot{U}(n,0)=0$ for the longitudinal displacements. Here random noise was chosen as a perturbation, not only because it is relevant to realistic experimental conditions but also since it is an efficient way to excite all the wave-numbers including the unstable ones. Moreover, we can confirm in this way our analytical results by identifying the two characteristic wave-numbers $K_c$ and $K_m$ using Eqs.(\ref{kc}) and (\ref{K_m}) during the lattice dynamics simulation.

Here we note the following technical point. Due to the periodic boundary conditions, the spectrum is wrapped between $[0;\pi]$. During the manifestation of the MI, we expect to identify at least the following wavenumbers: the carrier $k$, and the most unstable wavenumber $K_m$. However we know that we always excite at least the third harmonics $3k$. In order for all these frequencies to be well identified, we thus choose parameters such that the $k+\epsilon K_m$ is smaller than $3k$. 
To do so, we use an alternative representation of Fig.~\ref{Panels_delta_map}, using as colormap the values of $K_m$. The two points denoted by squares and circles in left and right panel respectively, are the two examples that we will study in details below.

\begin{figure}[h!]
   \centering
   \includegraphics[width=0.49\textwidth,trim = 0 0 0 0]{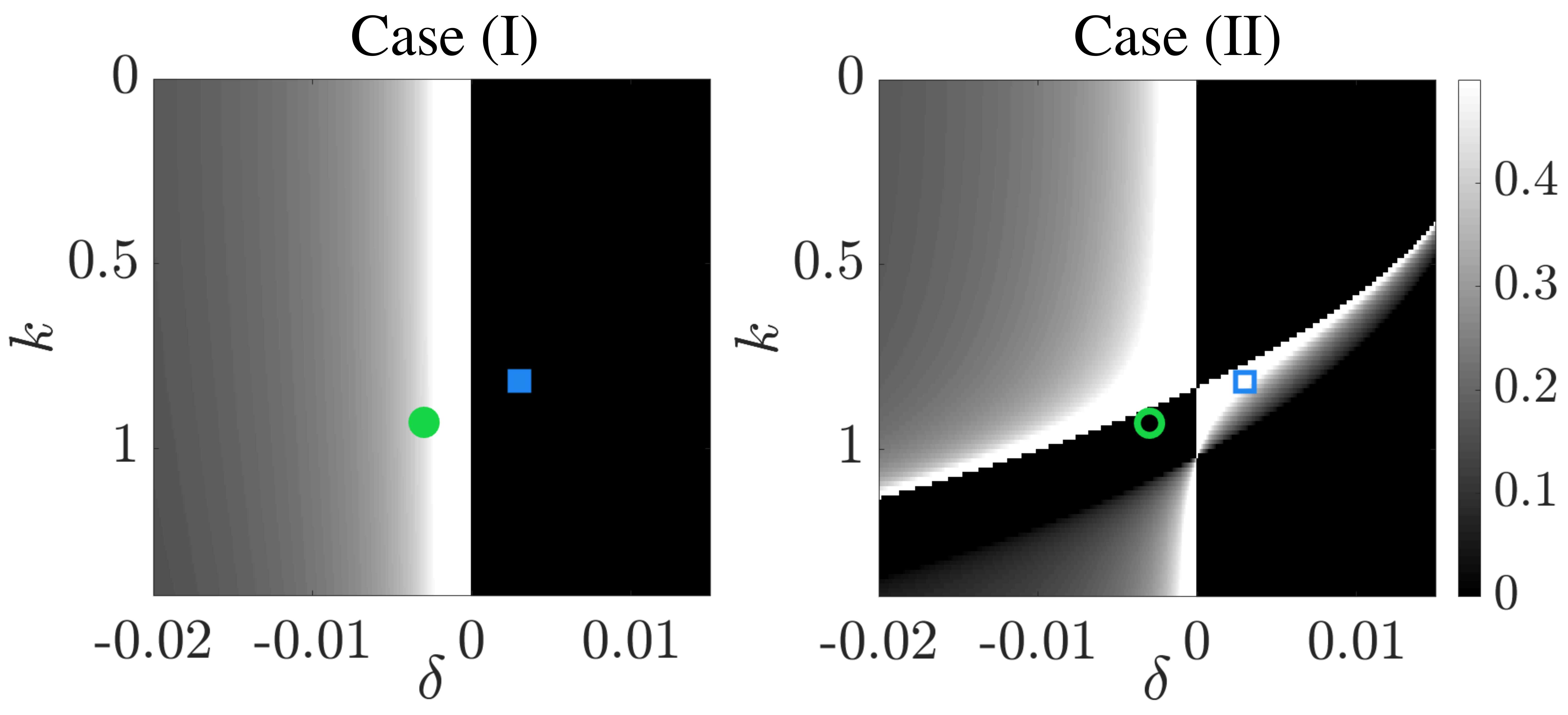}
  \caption{\label{K_inst_map} Most unstable wave number $K_m$ (colormap) as a function of $\delta$
 and $k$ for $\alpha=2.5$. 
 In both panels, two particular points are indicated: a blue square point for $k=0.81681$ and $\delta=0.003$ and a green circle point at the position $k=0.92991$ and $\delta=-0.003$.}
  \label{K_inst map}
\end{figure}

\subsection{Inducing MI by coupling the rotations with displacements}

We first focus on a point, in the parameter space spanned by $\delta$ and $k$, indicated by the square in Fig.~\ref{K_inst_map}. This corresponds to the plane wave wavenumber $k=0.81681$ and $\delta = 0.003$. We fix from now the value of $\alpha=2.5$. As a reminder, the values of $\delta$ fixes the difference between shear and bending stiffness ($\delta=K_s-K_\theta$), while the value of $\alpha$ the ratio of mass to moment of inertia of the particles. 

\begin{figure}[h!]
   \centering
   \includegraphics[width=0.49\textwidth,trim = 0 0 0 0]{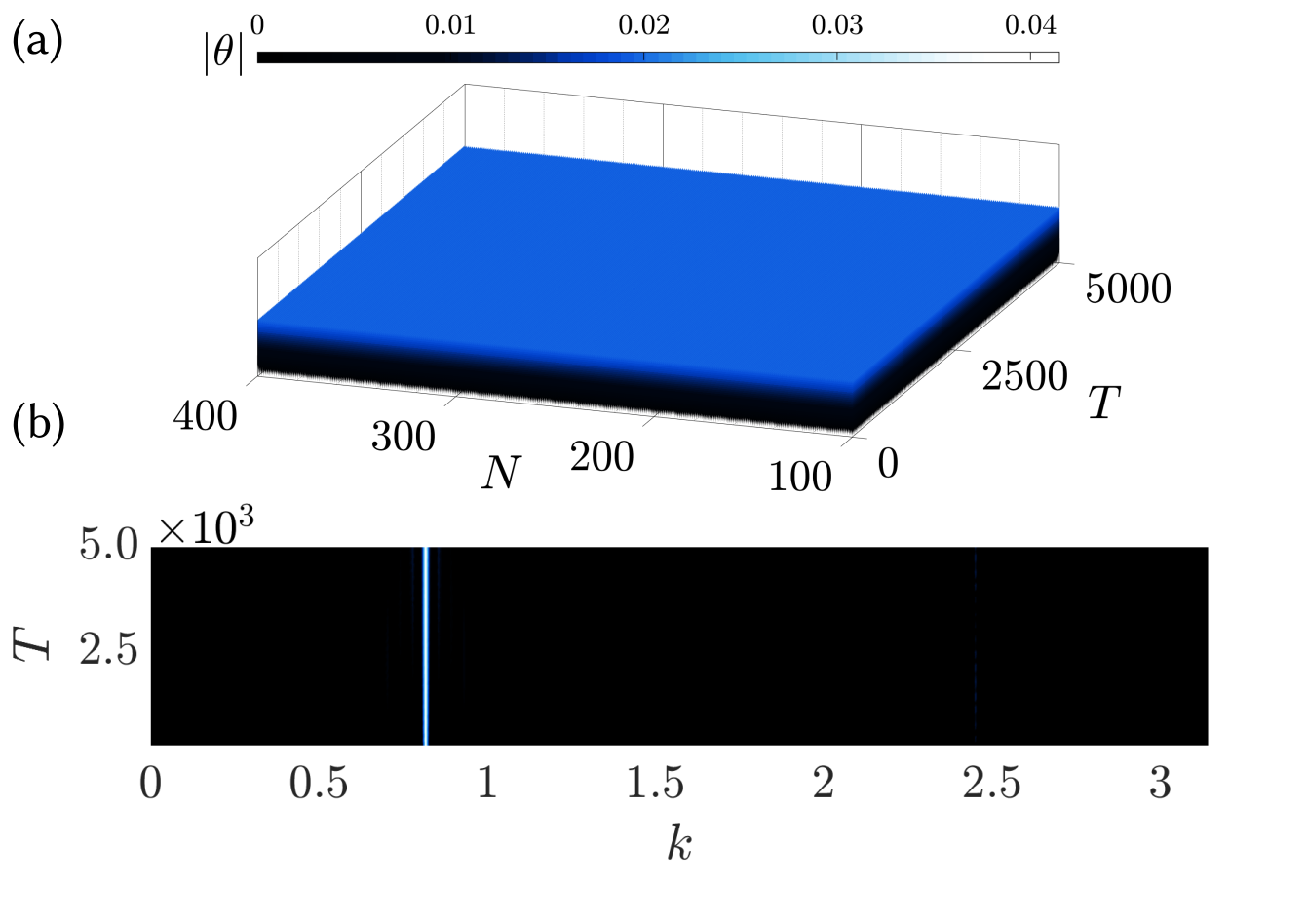}     
   \caption{\label{Blue_decoupled}
   Panel (a) represents the evolution in time ($T$) of the absolute value of the rotation amplitude along the chain ($N$). Panel (b) represents the evolution in time of its k-spectrum. The results correspond to the case (I), blue square point ($k=0.81681$, $\delta =0.003$) 
   } 
\end{figure}

According to the theory, this point is described by a defocusing NLSE ($g<0$) for the case (I). Thus, the plane wave is supposed to be stable. In contrary, when both the dof are considered, i.e., case (II), and for exactly the same parameters, the plane wave become modulationally unstable, since the system is described by a focusing NLSE ($g>0$).
To confirm our theoretical prediction, we solve the discrete system of Eqs.~(\ref{norm}) using the initial conditions (\ref{IC}).
In Fig.~\ref{Blue_decoupled}, we show the results for the case (I). Here both the evolution of the rotation [panel (a)] and its space Fourier transform [panel (b)] indeed show that a random perturbation on an initial plane wave remains bounded, thus the plane wave is stable. 

The rotations $\theta_n$ show small amplitude oscillations in time with a frequency $\omega$, following the dispersion relation Eq.~\eqref{continuum2} at the given $k$. Even after more than $500$ oscillations, only the wavenumber of the carrier wave is present in the spectrum, indicating the stability.


\begin{figure}[h!]
     \includegraphics[width=0.49\textwidth,trim = 0 0 0 0]{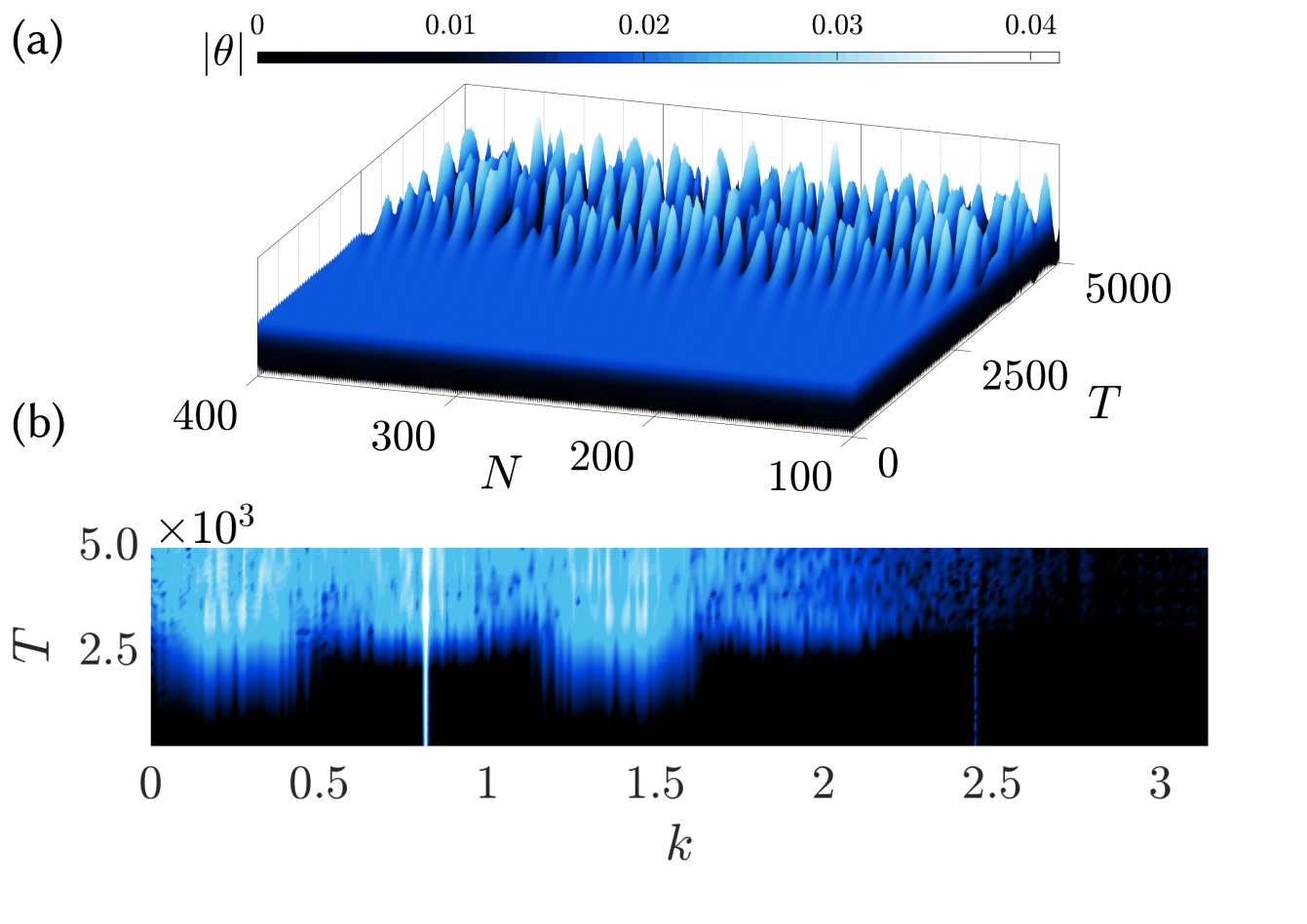} 
     \includegraphics[width=0.49\textwidth,trim = 0 0 0 0]{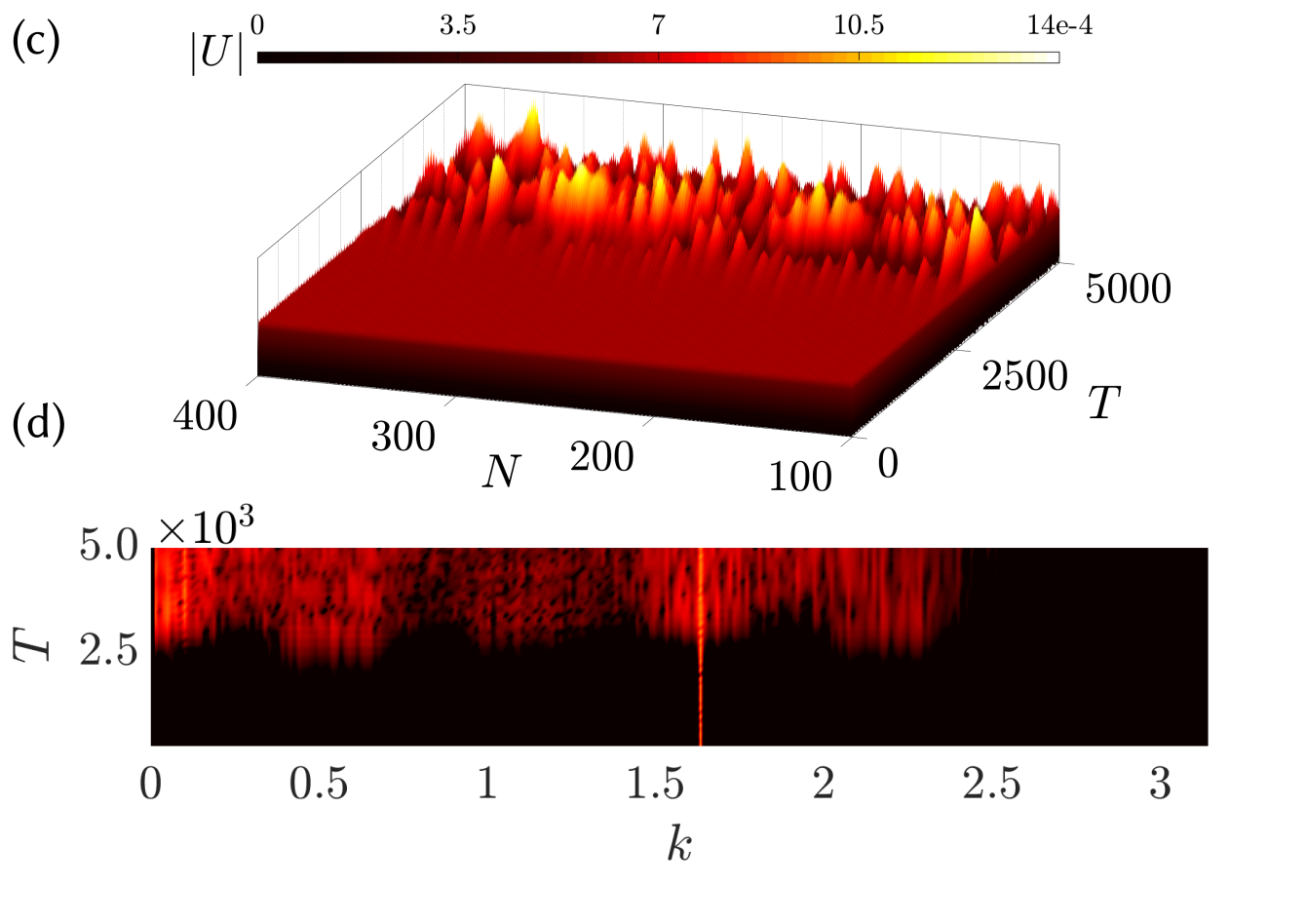}    
     \caption{\label{Blue_coupled}. 
     Panels (a-c) represent the evolution in time ($T$) of the absolute value of the rotation respectively displacement  amplitudes along the chain ($N$). Panels (b-d) represents the evolution in time of the k-spectrum for the rotation and longitudinal displacement. The results correspond to case (II) - blue square point ($k=0.81681$, $\delta =0.003$).}
\end{figure}

For the exact same flexMM and the same initial condition, if we allow the coupling between the two dofs, namely if we consider the case (II), the dynamics is radically different. This scenario is shown in Fig.~\ref{Blue_coupled}. As predicted by the theory, the wavenumbers of the perturbation that belong to the instability band, start growing. This is clear by the two sidebands that are
developed symmetrically around the excited wavenumber $k=0.81681$ in panel (b) of Fig.~\ref{Blue_coupled}.
More precisely, the center of these side bands corresponds to the point $k\pm \epsilon K_m$ since the most unstable wavenumber rises first. The generation of these wavenumbers is directly revealed on the rotations as large amplitude localized structures appear [see Fig.~\ref{Blue_coupled} (a)]. For later times, after the instability kicks in, and when the amplitude of the rotations becomes large enough, we observe a spectrum with many excited wavenumbers.

In this case, since rotations are coupled to the longitudinal displacements $U_n$, we expect to see some dynamics in the displacements too. Indeed, as expected from our analysis in Eq.~\eqref{u1}, $U$ starts oscillating with a wavenumber $2k$ as shown in Fig.~\ref{Blue_coupled}(d), and at later times following the evolution of $\theta$,  larger amplitude modulated waves are also emerging in the displacements $U_n$. 



\subsection{Stabilizing plane waves using the coupling of dofs}
\begin{figure}[h!]
 \includegraphics[width=0.49\textwidth,trim = 0 0 0 0]{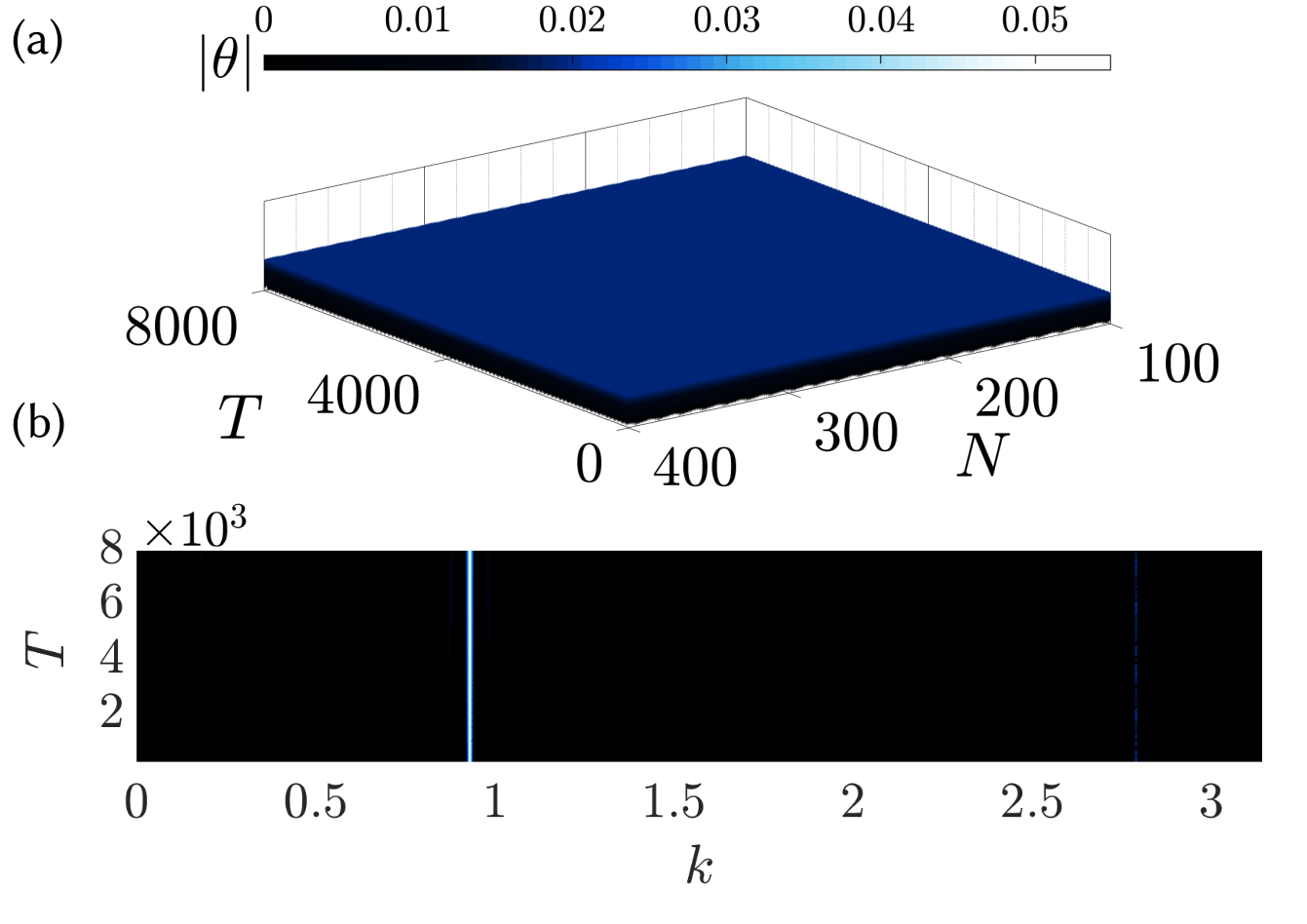} 
   \includegraphics[width=0.49\textwidth,trim = 0 0 0 0]{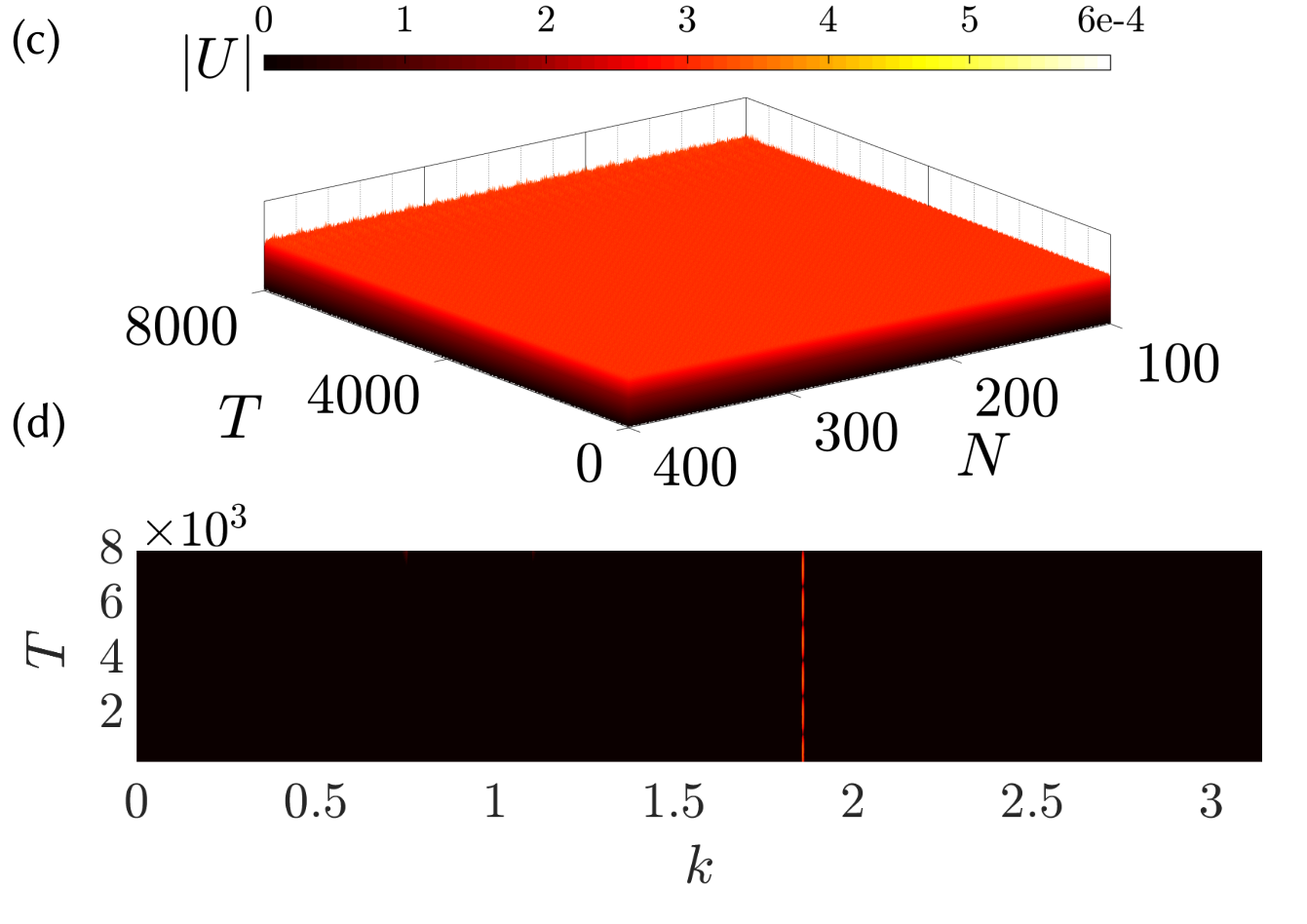}     
  \caption{\label{Green_coupled}Panels (a-c) show the evolution in time ($T$) of the absolute value of the rotation and displacement amplitudes along the chain ($N$). Panels (b-d) show the evolution in time of the k-spectrum for the rotation and longitudinal displacement. The results correspond to case (II), green circle point ($k =0.92991$, $\delta =-0.003$). }
\end{figure}

The second configuration which we focus on is the "complementary" one. It corresponds to the green circles in Fig.~\ref{K_inst_map}, where the uncoupled system (case I) is described by a focusing NLS, thus we expect the plane waves to be modulationally unstable, while by allowing the coupling between the two dofs (case II), the effective NLS is focusing and thus, the plane waves are stable.
\begin{figure}[h!]
   \centering
      \includegraphics[width=0.49\textwidth,trim = 0 0 0 0]{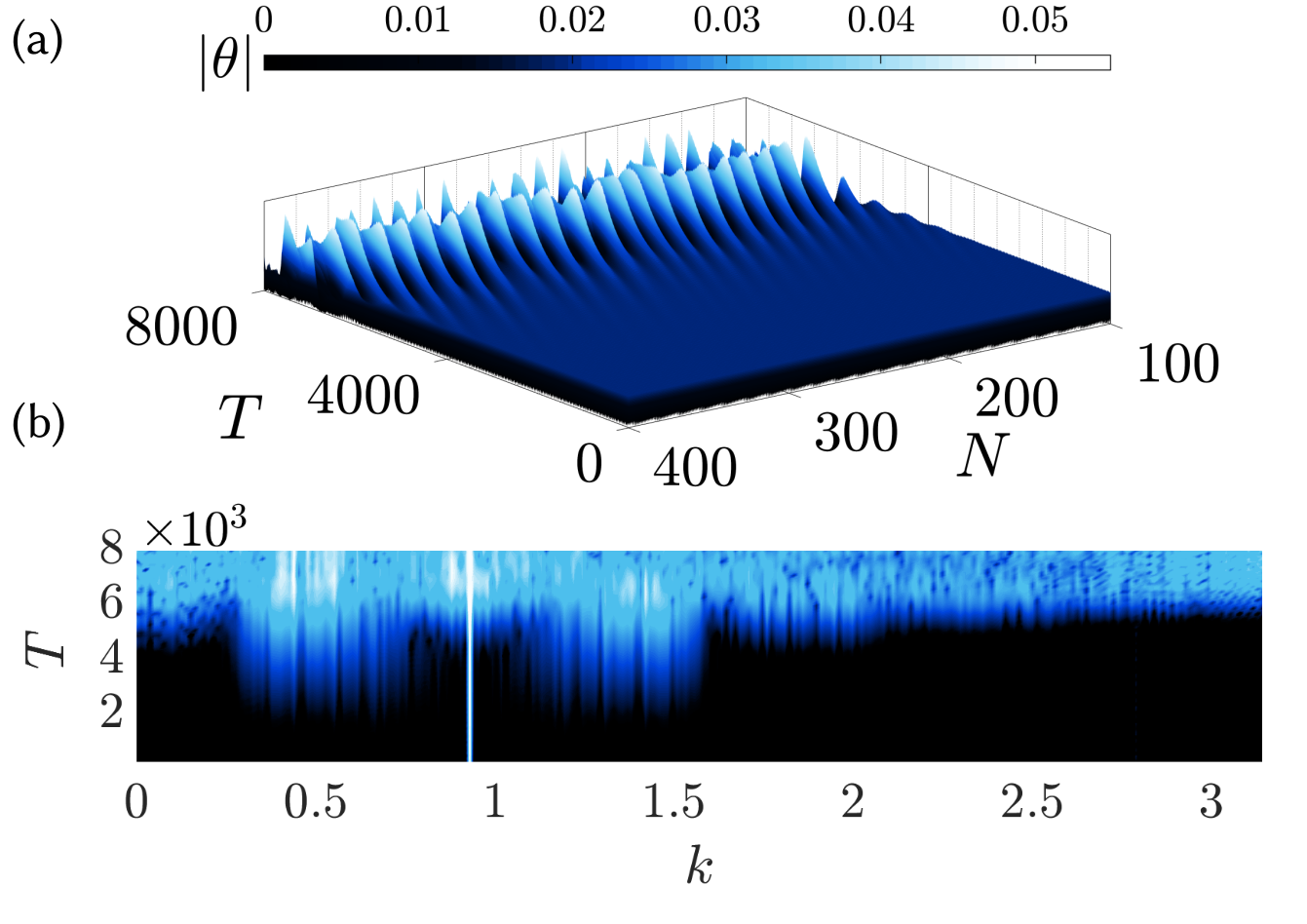}
\caption{\label{Green_decoupled}
 Panel (a) represents the evolution in time ($T$) of the absolute value of the rotation amplitude along the chain ($N$). Panel (b) represents the evolution in time of its k-spectrum. The result correspond to case (I), green circle point ($k =0.92991$, $\delta =-0.003$)}
\end{figure}
To confirm these theoretical predictions, we use the same initial conditions as in Eq.~(\ref{IC}) but with $k= 0.92991$ and $\delta=-0.003$ and we solve again numerically the system of Eqs.~(\ref{norm}).
The result of the case (I) is shown in Fig.~\ref{Green_decoupled}. Following our analysis, the numerical simulations confirm that an initially perturbed plane wave develops initially the expected side branches at $k\pm \epsilon K_m$. At the final steps of the simulation, all the wavenumbers are excited. On the other hand, when both dofs are present (case (II)) and for exactly the same parameter values, the corresponding numerical result, shown in Fig.~\ref{Green_coupled}(a-b), verifies the stability of the plane wave solution. We see that for the same total time of propagation as in the decoupled case, $\theta$ shows stable oscillations with a wavenumber $k$ while $U$ oscillates at $2k$, as per the theory.

\subsection{MI growth rate: theory vs numerics}
To further support our theoretical findings we perform numerical simulations, for both cases, by varying the amplitude of the initial excitation $A_0$, and we compare the predictions of the MI linear stability analysis [see Eqs.~(\ref{kc}-\ref{K_m})], with the early stage of the MI manifestation in simulations. 

\begin{figure*}
\centering
    \includegraphics[width=0.99\textwidth,trim = 0 0 0 0]{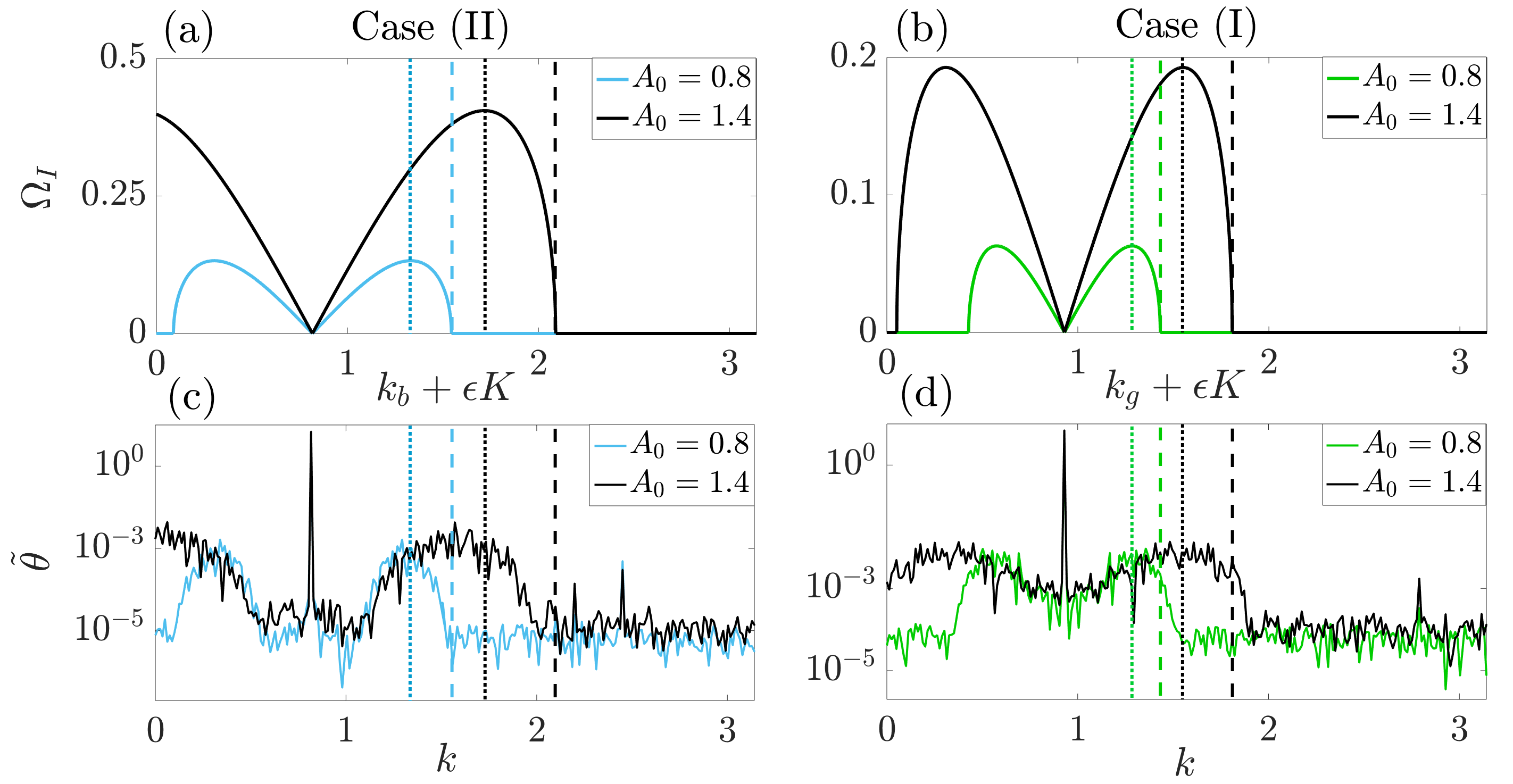}
  \caption{\label{Spectrumline}Panels (a) and (b) show the theoretical modulational instability band for the blue point in case (II), the green point in case (I) and for two different amplitudes $A_0$. In panels (c) and (d), we plot the Fourier transform of the $\theta$, corresponding to the parameters of the  blue square (respectively green circle) at two different instances. For case (II), we choose $t = 2500$ for $A_0=0.8$ and $t = 1100$ for $A_0=1.4$. For case (II), $t = 5000$ for $A_0=0.8$ and $t = 1700$ for $A_0=1.4$. The dotted lines correspond to the analytical values of $k_{b/g}+\epsilon K_m$, while the dashed ones to $k_{b/g}+\epsilon K_c$. $k_{b/g}$ are the wave numbers of the initial plane waves and $K$ the wave number of the perturbation.}
\end{figure*}

In particular, in Fig.~\ref{Spectrumline}(a-b), we plot the MI growth rate for both cases and for two different amplitudes $A_0$.  The dotted lines correspond to the analytical values of $k+\epsilon K_m$, while the dashed ones to $k+\epsilon K_c$.
In Fig.~\ref{Spectrumline}(c-d), we plot the Fourier spectrum of the rotation field $\theta$ at times that correspond to the early stage of MI.
In both cases a stronger initial excitation results in a larger unstable band showing a maximum shifted to larger $k$ values. Note also how well the theoretically predicted bandwidth, Fig.~\ref{Spectrumline}(a-b), matches the numerically obtained bandwidth. This observation constitutes another more quantitative validation of the derived effective NLS description.


\section{Conclusions}
FlexMM are mechanical structures with some unique features, including the geometrical nonlinearity coming from the large rotations of the building blocks, the presence of several dofs that are nonlinearly coupled, and the great tunability of the dispersion relation.
Therefore, flexMMs offer a perfect experimental platform to explore a plethora of nonlinear wave phenomena.

In this paper, we focused on the archetypal nonlinear phenomenon of MI. To that end, starting from a discrete, nonlinear lump model, that has been proved to accurately describe their dynamics, we first derived a NLS equation for slowly varying rotational envelope waves.
We then studied the stability of the rotational plane waves to small perturbations via the MI analysis for the derived NLS.
Analytical and numerical results revealed that, under proper values of the physical parameters of the flexMM,  namely under some particular values of the inertia and stiffness parameters, it is possible to observe MI in these flexMMs.
More importantly, we have analyzed the role of the coexistence of two dofs. 
In particular, the interplay between the two dofs can lead to regions of stability, in an otherwise unstable flexMM which supports only rotations, i.e., only one of the two dofs, and vice versa.

This work constitutes an attempt to understand more generally the dynamics of modulated waves in nonlinear flexMMs. Several natural extensions of this work include the initial excitation of both rotational and longitudinal modulated waves, leading to a coupled NLS with a much richer MI dynamics, as well as the study of the discreteness when shorter modulated waves are considered, along the lines of \cite{remoissenet_low-amplitude_1986,daumont_modulational_1997}. Both aspects are currently under investigation and results will be presented in future publications. Other interesting perspectives are the generation and dynamics of coherent structures like Peregrine breathers and extreme wave effects in nonlinear flexMMs.
We believe that the present work reveals the great potential that nonlinear flexMMs have, for the observation and control of both typical and novel nonlinear phenomena related to modulated waves. 

\section*{Acknowledgement}
The authors acknowledge the support from the project ExFLEM ANR-21-CE30-0003-01.
\newpage
\renewcommand\bibname{References}
\let\itshape\upshape
\normalem
\bibliography{Biblio_article.bib}

\appendix
\section{Dispersion relation}
The dispersion relation of the metastructure is obtained by linearizing the motion equations ($\sin\theta \approx\theta$) and assuming that the chain is excited by a harmonic source of $\omega$ pulsation, propagating along increasing x. The harmonic solutions of the linear system are represented by these three vectors, when one poses $x_i = ia$, $x_{i\pm1} = (i\pm 1)a$:
\begin{equation}
\begin{aligned}
    \vec{\phi}_{i}&=\left[\begin{array}{l}
U_{i} \\
\theta_{i}
\end{array}\right]=\left[\begin{array}{l}
U_{0} \\
\theta_0
\end{array}\right]e^{j(\omega t -kx_{i})}=\left[\begin{array}{l}
U_{i} \\
\theta_i
\end{array}\right],\\
  \vec{\phi}_{i+1}&=\left[\begin{array}{l}
U_{i+1} \\
\theta_{i+1}
\end{array}\right]=\left[\begin{array}{l}
U_{0} \\
\theta_0
\end{array}\right]e^{j(\omega t -kx_{i+1})}= \left[\begin{array}{l}
U_{i} \\
\theta_i
\end{array}\right]e^{-jka},\\
  \vec{\phi}_{i-1}&=\left[\begin{array}{l}
U_{i-1} \\
\theta_{i-1}
\end{array}\right]=\left[\begin{array}{l}
U_{0} \\
\theta_0
\end{array}\right]e^{j(\omega t -kx_{i-1})}=\left[\begin{array}{l}
U_{i} \\
\theta_i
\end{array}\right]e^{+jka}.
\end{aligned}
\label{solutionharm}
\end{equation}
By substituting these harmonic solutions in the linearized equations, we obtain the eigenvalue problem,

\begin{equation}
[M]^{-1}[K]\vec{\phi}=\lambda\vec{\phi}\; ,
\end{equation}

with $\lambda=\omega^2$ the eigenvalue and $\vec{\phi} =\left[\begin{array}{l}
U_{0} \\
\theta_{0}\\
\end{array}\right]$ the eigenvector, and
\begin{equation}
 \begin{aligned}
  \relax[M] &=  \begin{bmatrix} 1 & 0 \\ 0 & \alpha^{-2} \end{bmatrix} \, ,\\
     \\
[K] &=  \begin{bmatrix} 2\left(1-\cos(ka)\right) & 0 \\ 0& -2\delta\cos(ka)+2\left(K_s+2K_{\theta}\right) \end{bmatrix} \, .
 \end{aligned}
\end{equation}

The coupling between modes comes from the anti-diagonal terms of the $K$ matrix. Since all mass units are aligned at the initial time, these anti-diagonal coefficients are zero, so the modes are decoupled.

\section{Multiple scales}
\label{appendix:Multiplescales}
The different scales imply that the differentials of $X$ and $T$ must be redefined according to the different scales $X_i$ and $T_i$ used. By defining the notation $D_i = \frac{\partial}{\partial T_i}$ and in an analogous way $D_{iX} = \frac{\partial}{\partial X_i}$, we can write,

\begin{equation}
\begin{aligned}
    \frac{\partial^2}{\partial T^2} &= (D_0+\epsilon D_1 +\epsilon^2 D_2+...)^2 \\
    &= D_0^2+2\epsilon D_0D_1+\epsilon^2(D_1^2+2D_0D_2)+...\\
    \frac{\partial^2}{\partial X^2} &= (D_{0X}+\epsilon D_{1X} +\epsilon^2 D_{2X}+...)^2 \\
    &= D_{0X}^2+2\epsilon D_{0X}D_{1X}+\epsilon^2(D_{1X}^2+2D_{0X}D_{2X})+...
\end{aligned}
\label{diff}
\end{equation}

The operators $\mathcal{\hat{L}}_j^{(i)}$ and $\mathcal{\hat{M}}_j^{(i)}$ are given by the following expressions:

\begin{equation}
    \begin{split}
        \hat{\mathcal{L}}_0^{(1)} &= D_0^2-D_{0X}^2 \\
        \hat{\mathcal{L}}_0^{(2)} &= D_0^2-C_1 D_{0X}^2+C_2 \\
        \hat{\mathcal{L}}_1^{(1)} &=2(D_0D_1-D_{0X}D_{1X}) \\
        \hat{\mathcal{L}}_1^{(2)} &=2(D_0D_1-C_1D_{0X}D_{1X})\\
        \hat{\mathcal{L}}_2^{(1)} &=D_1^2-D_{1X}^2+2D_0D_2-2D_{0X}D_{2X}\\
        \hat{\mathcal{L}}_2^{(2)} &=D_1^2-C_1D_{1X}^2+2D_0D_2-2C_1D_{0X}D_{2X}\\
         \\
         \hat{\mathcal{M}}_0^{(1)} &= \frac{1}{2}D_{0X} \\
         \hat{\mathcal{M}}_0^{(2)} &=  -C_4D_{0X} \\
         \hat{\mathcal{M}}_1^{(1)} &= \frac{1}{2} D_{1X} \\
         \hat{\mathcal{M}}_1^{(2)} &=  -C_4D_{1X} \\
         \hat{\mathcal{M}}^{(3)} &=  -C_3 \\
    \end{split}
    \label{operators}
\end{equation}
\end{document}